\documentclass[aps,prl,twocolumn]{revtex4}

\newcommand{\be}{\begin{equation}}
\newcommand{\ee}{\end{equation}}

\begin{document}

\large

\title{Group-Theoretical Derivation of Angular Momentum Eigenvalues in
Spaces of Arbitrary Dimensions}

\author{Tamar Friedmann and C. R. Hagen\footnote{mail to:
tamarf@pas.rochester.edu, hagen@pas.rochester.edu} }
\affiliation{Department of Physics and Astronomy\\
University of Rochester\\
Rochester, N.Y. 14627-0171}

\begin{abstract}

The spectrum of the square of the angular momentum in arbitrary dimensions 
is derived using only group theoretical techniques.  This is accomplished by
application of the Lie algebra of the noncompact group $O(2,1)$.

\end{abstract}



\maketitle


The famously soluble problems of quantum mechanics include the simple harmonic oscillator, the hydrogen atom, and angular momentum.  Each of these is known to have solutions obtainable from careful analysis of the underlying wave equations as well as from purely algebraic methods.  In the case of the oscillator, the algebraic method is arguably preferable while for the hydrogen atom the choice between a wave equation approach and an algebraic one based on the Lenz vector may be less compelling.  Nonetheless, it is generally recognized that the two approaches constitute a valuable complement to each other.  In the case of angular momentum, Louck {\it et al.} \cite{Louck} have shown that the eigenvalues of angular momentum squared in $q$ spatial dimensions (namely, $\ell(\ell+ q-2)$) can be obtained equally well (if tediously!) from the relevant differential equations or from the underlying algebra. It is the object of the present note to demonstrate that these eigenvalues follow in a remarkably simple way from a third method based on group theoretical considerations.

The technique employed is based on the Lie algebra of $O(2,1)$.  This
particular algebra was invoked some decades ago by Bacry and Richard \cite{Bacry} to
obtain the spectra of the $q$-dimensional oscillator and the relativistic
hydrogen atom \cite{atom}.  Their techniques were in a sense a hybrid using the tools of
both differential equations as well as group theory.  The techniques presented
here, however,  are based solely on group theory, and they allow a totally group theoretical derivation when applied to the cases considered in \cite{Bacry}. 
The absence of such a result in the published literature constitutes a gap which this note proposes to address.

 One begins by
noting that the coordinate operators $x_i$ and $p_i$, where $i=1,2,...q$
with
$[x_i,p_j] = i \delta_{ij}$ and $\hbar=1$, may be expressed in terms of the
operators $a_i$ and $a^\dagger_i$ by
$$x_i=(a_i+a^\dagger_i)/ \sqrt{2}  $$
and
$$p_i = (a_i - a^\dagger_i)/ \sqrt{2} i$$
where $[a_i,a^\dagger_j] = \delta_{ij}$.  One readily infers that
$$x_ip_j-x_jp_i=-i[a^\dagger_ia_j-a^\dagger_ja_i]$$
so that the square of the angular momentum operator $L_q^2$, as defined by
\[ L_q^2={1\over 2}\sum_{i,j=1}^q L_{ij}^2~,\]
where $L_{ij}\equiv x_ip_j-x_jp_i$, becomes
\[ L_q^2=N_q(N_q+q-2)-A_q^\dagger A_q , \]
where  $N_q\equiv a^\dagger_ia_i, A_q\equiv a_ia_i$.
The operators $N_q$ and $A_q$ satisfy the commutation relations
$$[N_q,A_q]=-2A_q$$ 
and
$$[A_q,A_q^\dagger]=4N_q+2q.$$

Upon defining $K_{q-} = {1\over 2}A_q$, $K_{q+}= {1\over 2}A_q^\dagger$, $K_{q1}= {1\over
2}(K_{q+}+K_{q-})$,   $K_{q2}= {1\over 2i}(K_{q+} -K_{q-})$, and $J_{q3} = {1\over 2}N_q+{q\over
4}$, one obtains the $\mathfrak{o}(2,1)\sim  \mathfrak {su}(1,1)$ Lie algebra
\cite{algebra}
$$[J_{q3},K_{q1}]=iK_{q2},$$
$$[J_{q3},K_{q2}]=-iK_{q1},$$
$$[K_{q1},K_{q2}]=-iJ_{q3}.$$
The Casimir operator
$$Q _q\equiv J_{q3}^2-K_{q1}^2-K_{q2}^2$$
is readily seen to commute with each of the individual members of the
algebra $J_{q3}$, $K_{q1}$, and $K_{q2}$.  
Comparison with the result for $L_q^2$ yields
the relation
\be \label{l2q} L_q^2 = 4Q_q-{1\over 4} q^2+q, \ee
so that the spectrum of $L_q^2$ is determined by values of the Casimir operator
of representations of  $\mathfrak{o}(2,1)$ \cite{Jackiw}.  Generalizing the earlier work
of Bargmann \cite{Bargmann} (who considered the group $O(2,1)$, and not its covering
group ${\overline{SU(1,1)} } $ ), these values have been obtained 
via algebraic techniques
by Barut and
Fronsdal \cite{BF} who showed that there are four series of representations of 
$\mathfrak{o}(2,1)$ characterized by a pair of complex numbers $(\Phi,
E_0)$. The series are
\begin{enumerate}
\item $D(\Phi, E_0)$, where $-{1\over 2}<Re(E_0)\leq {1\over 2}$, $\Phi \pm
E_0\neq \pm n$ ($n$ is a non-negative integer) and the $J_{q3}$ spectrum is
$E_0\pm n$;
\item $D^+(\Phi)$, where $E_0=-\Phi$, $2\Phi\neq n$, and  the $J_{q3}$ spectrum
is $E_0+n$; \item $D^-(\Phi)$, where $E_0=\Phi$, $2\Phi \neq n$, and the
$J_{q3}$ spectrum is $E_0-n$; and \item $D(\Phi)$, where $E_0=0$, $2\Phi = n$,
and the $J_{q3}$ spectrum is $-\Phi, -\Phi +1, \ldots , \Phi -1, \Phi$.
\end{enumerate}
Additional conditions on $\Phi$ and $E_0$ are needed for these representations
to be unitary: $D(\Phi, E_0)$ is unitary if either $Im (E_0)=0$ and $\Phi =
-{1\over 2}+i\lambda$ (``principal series") or $Im (E_0)=Im (\Phi) = 0$ and
$|\Phi +{1\over 2}|<{1\over 2}-|E_0|$ (``supplementary series"); $D^+(\Phi)$
and $D^-(\Phi)$ are unitary if $Im (E_0)=0$ and $\Phi <0$; and $D(\Phi)$ is
unitary only for the trivial representation $\Phi=0$. Also note that there are no degeneracies in the $J_{q3}$ spectra.

For all the above representations, the Casimir operator is given by
\be Q_q=\Phi (\Phi +1) .\ee
Since $J_{q3} = {1\over 2}N_q+{q\over 4}$ and the number operator $N_q$ is a
non-negative integer \cite{integer}, one knows that $J_{q3}$ is bounded below and positive. Only
$D^+(\Phi)$ with $E_0>0$ satisfies this condition. Since
\[ E_0=-\Phi ,\]
one has  $\Phi <0$ so the representation is unitary, and the  spectrum of $J_{q3}=
{1\over 2}N_q+{q\over 4}$ is $E_0, E_0+1, E_0+2, \ldots $. 

Thus the spectrum of $L_q^2$ consists at most of the values obtained from
equation (1) by inserting $Q_q=\Phi (\Phi +1)=E_0(E_0-1)$ for $E_0=
{1\over 2}\ell_q+{q\over 4}$ for some integer $\ell_q \geq 0$.
It remains to be shown that all integers $\ell_q\geq 0$ are obtained.

Rewriting the Casimir operator as
\[ Q_q= J_{q3}(J_{q3}-1)-K_{q+}K_{q-} \]
shows that if one finds a state $|\psi \rangle$ in a representation of
$\mathfrak{o}(2,1)$ that is annihilated by $K_{q_-}$, that state will satisfy
\be \label{QJ3} Q_q|\psi \rangle=J_{q3}(J_{q3}-1)|\psi \rangle ,\ee
so that the state's $J_{q3}$ eigenvalue will equal the $E_0$ of that representation. It
may also be noted that the operator $K_{q_-}$ plays the role of a lowering
operator in the sense that it lowers the eigenvalues of $J_{q3}$ by one unit.
One now proceeds to obtain such a state $|\psi _{\ell _q}\rangle$ for each $\ell_q
\geq 0$.

  To this end one considers the Hilbert space of number eigenstates
$$| \{n_i\} \rangle =|n_q, n_{q-1}, \ldots , n_1\rangle$$
on which the operators $N_{qj}\equiv a_j^\dagger a_j$ (no sum), $a_i$, and $a_i^\dagger$ act via
$$ N_{qj}| \{n_i\} \rangle = n_j| \{ n_i \}  \rangle ;$$
$$a_j| \{n_i\} \rangle =  \sqrt{n_j} | \{n_i-\delta_{ij}\} \rangle ; $$
$$ a_j^\dagger| \{n_i\} \rangle =  \sqrt{n_j+1} | \{n_i+\delta_{ij}\} \rangle.$$ 
One then defines operators
$$a_{\pm}={1\over {\sqrt 2}}(a_2\pm ia_1)$$
so that $[a_{\pm},a_{\pm}^{\dagger}]=1$ is the only nonvanishing commutator in 
the set of $a_{\pm}$ and $a_{\pm}^{\dagger}$.  
Writing
$$K_{q-}=a_+a_- +{1\over2}\sum_{i=3}^q a_ia_i$$
it follows that
$$[K_{q-},(a_{\pm}^{\dagger})^{\ell_q}]=\ell_q(a_{\pm}^{\dagger})^{\ell_{q-1}} a_{\mp}.$$
Thus an appropriate state $|\psi_{\ell_q}\rangle$ which 
vanishes when acted on by $K_{q-}$ is easily seen to be
\be \label{psil} | \psi_{\ell_q} \rangle = (a_{\pm}^{\dagger})^{\ell_{q} }| \psi _0\rangle\ee 
where $N_q |\psi_0\rangle =0$.
This establishes that \emph{all} eigenvalues of $J_{q3}$ are possible $E_0$'s
and upon insertion into the relations  (\ref{l2q}) and (\ref{QJ3})  one sees
that the eigenvalues of $L_q^2$ are given by
$$L_q^2=\ell_q(\ell_q +q-2),$$
where $\ell_q$ is any non-negative integer. Thus the eigenvalue spectrum of
$L_q^2$ follows from strictly group
theoretical considerations.

The extension to the additional members of the 
complete set of commuting operators 
(CSCO) 
$L_q^2,L_{q-1}^2, \ldots , L_3^2,L_{12}$ is now immediate. 
One notes first that the state (\ref{psil}) is is one of \cite{dims}
\be \label{dim} (q+2\ell _q-2)\frac{(q+\ell _q-3)!}{\ell_q ! (q-2)!} \ee
states satisfying $N_q|\psi \rangle = \ell _q|\psi \rangle $ and $K_{q-}|\psi \rangle = 0$ and forming an irreducible representation of $SO(q)$.
For each value of $q'<q$ one defines operators $K_{q'\pm}$,
 $J_{q'3}$, and $N_{q'}$. 
 One readily infers the eigenvalues of $L_{q'}^2$, which are Casimir operators of $SO(q')\subset SO(q)$, to be of the form $\ell_{q'}(\ell_{q'}+q'-2)$ where the $\ell_{q'}$ are integers satisfying the relations\be \label{ells}\ell_q\geq\ell_{q-1}\geq\ell_{q-2}\geq \cdots \geq \ell_2\geq 0 .\ee
One deduces from the branching rules \cite{branch} of the above representation  that all possible combinations of $\ell _{q'}$ are attained.

Since this procedure yields only the eigenvalues of the square of $L_{12}$ despite the fact that the \emph{sign} of $L_{12}$ is to be included in the commuting set of operators, some additional comment is warranted in that case. To this end one notes that
$$L_{12}=a_+^{\dagger}a_+ -a_-^{\dagger}a_-$$
which anticommutes with the parity operator $P$.  The latter has the property that
$$Pa_1P^{-1}=-a_1 $$
while leaving all $a_i$ unchanged for ${i>1}$.  Since $P$ commutes with the operators $L_q^2,L_{q-1}^2,...L_2^2$, a CSCO necessarily requires that $P$ be added to this set.  Thus all states with nonzero $\ell_2$
are twofold degenerate.  In this two-dimensional space one can readily diagonalize the operator $L_{12}$ to obtain eigenvalues $\pm\ell_2$.
 This completes the determination of the full set of
eigenvalues of the operators $L^2_q,L^2_{q-1},\ldots , L^2_3, L_{12}$ \cite{12}.

\vskip .3cm
\noindent {\bf Acknowledgments}: Illuminating discussions with Jonathan Pakianathan and his comments on a draft are gratefully acknowledged. This work was supported in part by US DOE Grant number DE-FG02-91ER40685. 


\begin{thebibliography}{99}
\bibitem{algebra}
Following standard usage one denotes the group and its associated Lie algebra by upper and lower cases respectively.
\bibitem{integer}
It is worth noting that the nonnegative integer spectrum of $N_q$ can be obtained directly from the $J_{3}$ spectrum of suitable representations of $o(2,1)$ by applying
the group theoretical approach of \cite{Bacry} to a sum of $q$ one-dimensional oscillators.  No recourse to analytic or algebraic methods is required.
\bibitem{dims}
The result (\ref{dim}) is obtained by subtracting the dimension of the space with $N_q=\ell _q-2$ from the dimension of the space with $N_q=\ell _q$.
\end{thebibliography}
\end{document}